\documentclass[]{spie}  
\usepackage[]{graphicx}
\usepackage{amssymb,amsmath}
\title{The Vector-APP:\\a Broadband Apodizing Phase Plate that yields Complementary PSFs} 

\author{Frans Snik\supit{a}, Gilles Otten\supit{a}, Matthew Kenworthy\supit{a}, Matthew Miskiewicz\supit{b}, Michael Escuti\supit{b},  
Christopher Packham\supit{c}, Johanan Codona\supit{d}\skiplinehalf
\small{\supit{a}Sterrewacht Leiden, Universiteit Leiden, Niels Bohrweg 2, 2333 CA Leiden, the Netherlands; \\
\supit{b}Department of Electrical and Computer Engineering, North Carolina State University, Raleigh, NC 27695, USA; \\
\supit{c}Department of Astronomy, University of Florida, 211 Bryant Space Science Center, Gainesville, FL 32611, USA; \\
\supit{d}Steward Observatory, University of Arizona, 933 North Cherry Avenue, Tucson, AZ 85721, USA.}
}

\authorinfo{email: {\tt snik@strw.leidenuniv.nl}}

\begin{document} 
  \maketitle 

\begin{abstract}
The apodizing phase plate (APP) is a solid-state pupil optic that clears out a D-shaped area next to the core of the ensuing PSF.
To make the APP more efficient for high-contrast imaging, its bandwidth should be as large as possible, and the location of the D-shaped area should be easily swapped to the other side of the PSF.
We present the design of a broadband APP that yields two PSFs that have the opposite sides cleared out.
Both properties are enabled by a half-wave liquid crystal layer, for which the local fast axis orientation over the pupil is forced to follow the required phase structure.
For each of the two circular polarization states, the required phase apodization is thus obtained, and, moreover, the PSFs after a quarter-wave plate and a polarizing beam-splitter are complementary due to the antisymmetric nature of the phase apodization.
The device can be achromatized in the same way as half-wave plates of the Pancharatnam type or by layering self-aligning twisted liquid crystals to form a monolithic film called a multi-twist retarder.
As the VAPP introduces a known phase diversity between the two PSFs, they may be used directly for wavefront sensing.
By applying an additional quarter-wave plate in front, the device also acts as a regular polarizing beam-splitter, which therefore furnishes high-contrast polarimetric imaging. 
If the PSF core is not saturated, the polarimetric dual-beam correction can also be applied to polarized circumstellar structure.
The prototype results show the viability of the vector-APP concept.
\end{abstract}



\section{CORONAGRAPHY WITH APODIZING PHASE PLATES}

The direct detection of extrasolar planets enables insight into the configuration and architecture of planetary systems around nearby stars in our galaxy, testing current theories and mechanisms of formation.
Analysis of the light of an exoplanet enables the unambiguous characterization of the planetary atmosphere and surface, and ultimately allows for the detection of biomarkers (e.g. the presence of oxygen and liquid water) at rocky planets.
Direct imaging is primarily limited by the proximity of the parent star, where the light from the planet can be several decades fainter than the extended halo of light diffracted by the telescope and science camera optics. 
In order to minimize the contrast between star and planet, several methods are used together. 
Adaptive optics (AO) systems increase the encircled energy of the planets signal whilst reducing the diffracted host star light. 
Using models of the telescope point spread function (PSF) are limited by the presence of time varying optical aberration errors that are unsensed by the telescope's AO system, adding an additional component of noise that does not decrease as $\sqrt{t}$ with the total observation time. 
Observing at longer wavelengths towards the black body peak emission of the planet at infrared and thermal wavelengths reduces the contrast by a factor of a hundred to a thousand.
Polarimetric observations allows one to reduce the contrast between the unpolarized starlight and the polarized light of the scattered light off planet by several order of magnitude at visible wavelengths.

In addition to these  techniques, coronagraphs are used to suppress the diffraction halo from the central star whilst letting the flux from the planet through. Coronagraphs\cite{coronagraphyintro0, coronagraphyintro1, coronagraphyintro2, coronagraphyintro3} represent a tradeoff between planet flux throughput, angular resolution, azimuthal range of effectiveness and the smallest angle at which the coronagraph provides suppression.
A study by Guyon (2006)\cite{coronagraphyintro1} investigates the diverse range of theoretical coronagraph designs and their effectiveness, demonstrating that several coronagraphic designs are additionally limited by the finite angular size of the star itself, an importantconsideration for the closest stars to the solar system.

The Apodizing Phase Plate (APP \cite{APP1,APP2,APP3,APP4}) is a transmissive optic that modifies the phase of incoming incident wavefront at a pupil plane in the science camera. 
The point spread function (PSF) of the science camera is modified for all objects in the field of view, suppressing diffraction over a 180 degree wedge surrounding all objects. 
The current designs have suppression from 2 to 7$\lambda/D$ and have typically 2 to 4 magnitudes of suppression to give $10^{-4}$ contrast with respect to the peak flux. Several advantages include: 
\begin{itemize}
\item[+] Only one optic in the pupil plane;
\item[+] Insensitive to tip-tilt errors in the telescope and AO system;
\item[+] Consistent throughput for the planet at all angles;
\item[+] Small inner working angle;
\item[+] No reduction of the effective pupil size $D$.
\end{itemize}
Limitations to the current APP designs include:
\begin{itemize}
\item[--] The 180 degree coverage;
\item[--] The chromaticity of the current realised designs.
\end{itemize}

The APP is chromatic as it suffers from a $1/\lambda$ term in the locally applied phase, and chromatic terms due to dispersion of the substrate glass are present as well. 
In practice, therefore, it takes two observations (and thus often two nights) to observe the complete circumstellar environment of a target, as the APP needs to be physically rotated. 
In any case, the two PSFs can never obtained under identical conditions.
Nevertheless, interesting science results have already been obtained using the APP coronagraph.\cite{NaCoAPP1,NaCoAPP2}

With the vector-APP, we aim to solve the issues of the 180 degree coverage and chromatism.
The first issue is mitigated by producing two PSFs that have cleared-out regions on either side of the PSF core.
The second issue is solved by applying the required phase pattern through the vector phase that is determined by the orientation of a half-wave liquid crystal device.

\section{VECTOR PHASE APODIZATION}
Most optical components like lenses and nonplanar mirrors induce a phase pattern upon the incident light beam.
Also many coronagraphs aim to move away and/or null out the light of the central star through a phase manipulation.

Phase patterns are often induced through shape variations upon an isotropic substrate, but they can also be created by a plane-parallel birefringent medium, through the so-called vectorial phase or Pancharatnam\cite{Pancharatnamphase}-Berry\cite{Berryphase} phase.
Ideally, a vector-phase device is a half-wave plate, which fast axis orientation is variable.
To understand how vector phases are introduced by half-wave retarders, one should consider the incident light to be composed of two orthogonal circular polarization states: right- and left-handed.
Regardless of the orientation of a half-wave retarder, it flips the handedness of circular polarization: left- into right-handed circular polarization, and vice versa.
The (relative) phase of the emergent light is directly controlled by the orientation of the half-wave retarder: a rotation of $\theta = 0$---$\pi$ of the fast axis yields a phase variation in the emergent light of $\phi_R = 2 \theta = 0$---$2\pi$ for the right-handed circular polarization state.
For the other polarization state, the induced phase is negative ($\phi_L = -2 \theta$) for the same (positive) rotation $\theta$ of the half-wave retarder.
This way, a full wave of phase delay can be imposed by a 180$^\circ$ rotation of the half-wave retarder's fast axis.
A true and unhindered wraparound in phase (i.e., phase variation beyond a 2$\pi$ range) is thus easily created by continuously rotating the axis, contrary to a Fresnel lens, where ramps are required to effectuate a phase wrap.

To create a phase \emph{pattern} through the vector phase, a half-wave retarder with a specific pattern to the orientation of its axes is required. 
Such a device is enabled by liquid crystal (LC) and photo-alignment technology\cite{Gibbons}, developed for the LC display industry\cite{Schadt1, Broer} and patterned retarder optics\cite{Schadt2, Crawford, Marrucci, McEldowney, vectorvortextech}.
The orientation pattern is inscribed in a layer consisting of a UV-sensitive photoalignment material, which aligns itself along the linear polarization direction of the UV light. 
A birefringent LC layer, often polymerizeable\cite{Broer}, is deposited in one or more coatings on top of this alignment layer, and orients itself along the prescribed direction pattern. 
It is a homogenous retarder in the local neighborhood, so the thickness must be chosen such that it becomes half-wave at a certain wavelength.
Several methods exist to create an achromatic half-wave retardance (see Sect.~\ref{achr}), many of which employ LCs. 
The most common approach\cite{Pancharatnam} results in an element that can be a solid part, formed by laminating several initially separate elements, and electronically switchable layers may be added. 
The typical accuracy of the orientation pattern in this approach is \hbox{$\sim$1 $\mu$m}, outside of the important regions where the pattern orientation changes rapidly\cite{Mawet1}. 
In a newly developed approach\cite{Komanduri+2012}, multiple layers of twisted LCs are arranged into a monolithic film as a single element from the start, on a single substrate and single alignment layer. 
Any retardation spectra can be easily designed, including achromatic quarter- and half-wave plates at wide and ultra-wide bandwidths\cite{Komanduri2}.
Commonly used LC materials are transparent over large portions of the spectrum from 0.4--40 $\mu$m, with some typical hydrocarbon resonant absorption features between 6 and 11 $\mu$m\cite{polarizationgrating, Wu}.
Specialized materials with customized chemistry may reduce or shift these absorption bands.

The vector phase principle in this context has been applied to the four-quadrant phase mask (4QPM\cite{4QPM1,4QPM2,4QPM3}) and to the vortex coronagraph\cite{Mawet1, Mawet2, Mawet3, Mawet4, Mawet5}.
Both these coronagraph masks are to be inserted in an image plane, where on the optical axis they null out the light of the star.
The diffracted light from the mask can is mostly located outside the next pupil image, where it can be blocked by a Lyot mask.
The vector vortex coronagraph has a theoretically excellent performance.
Liquid crystal technology permits to manufacture vortex phases of nearly any topological charge, without any instantaneous phase jumps from 2$\pi$ back to 0.
Furthermore, the device can be made achromatic (see below).
One significant disadvantage of the vector vortex coronagraph is that it relies heavily on very precise tip/tilt correction of any guiding offset or other drift.
A small tip/tilt error already leads to a strong leak of starlight.
The APP, as it is inserted into a pupil plane, is completely insensitive to such tip/tilt errors, which makes it a very practical coronagraph.
By turning it into a vector-APP, we significantly enhance the APP's performance.

\begin{figure}[t]
\centering
\includegraphics{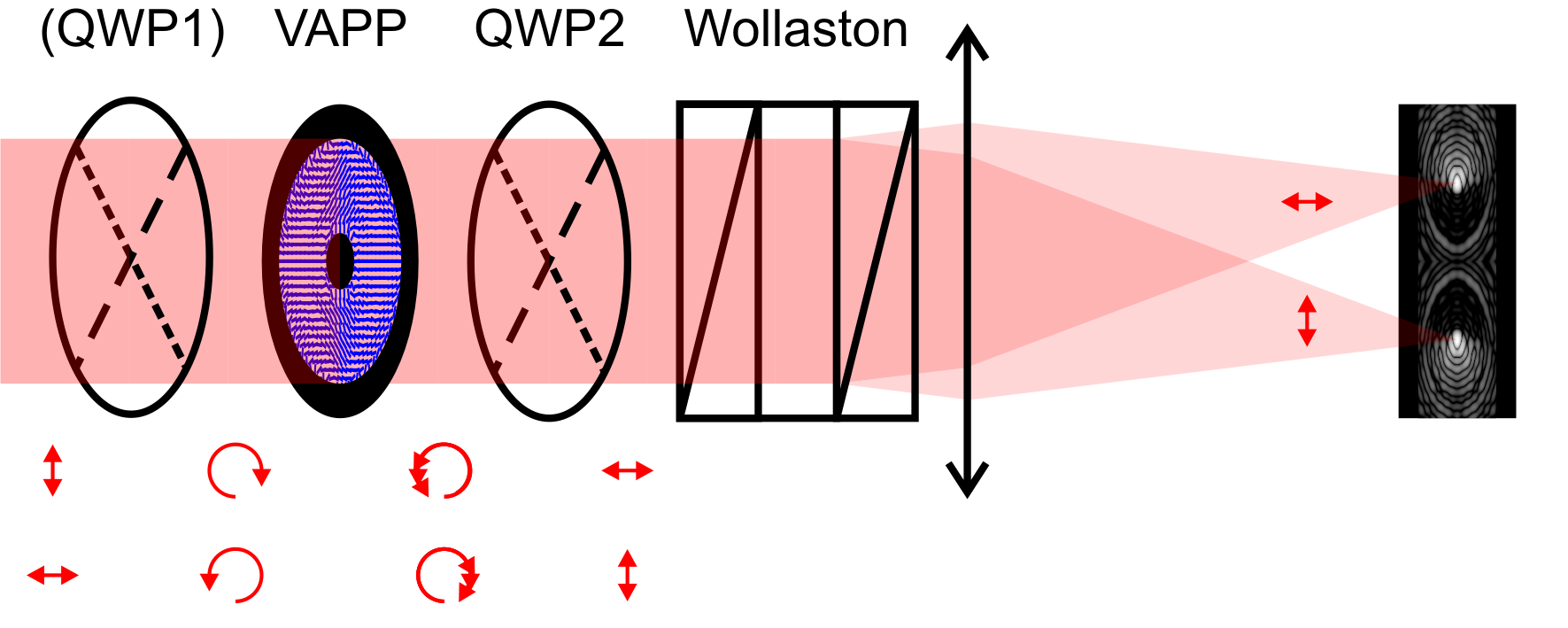}
\caption{Working principle of the vector-APP.}
\label{VAPPprinciple}
\end{figure}

\newpage
\section{THE VECTOR-APP}

The vector-APP is a LC half-wave device that has the phase pattern $\phi(x,y)$ of the APP inscribed into the axis orientation $\theta(x,y)=\phi(x,y)/2$ (mod $\pi$) of the liquid crystals.
The resulting vector phases $\phi_R(x,y)$, $\phi_L(x,y)$ are opposite ($\phi_R(x,y) = -\phi_L(x,y)$).
This phase pattern $\phi(x,y)$ is antisymmetric (in $x$), and therefore the resulting vector phases are antisymmetric.
This implies that the two PSFs belonging to the two circular polarization states are antisymmetric:
\begin{equation}
\mathcal{F}(\phi_R(x,y)) = \mathcal{F}(\phi_L(-x,y))\,,
\end{equation}
with $\mathcal{F}()$ denoting the Fourier transform.
This means that these two PSFs are complementary in the sense that they have the opposite sides of the PSF cleared out and their D-shaped dark holes together cover the complete 360$^\circ$ of space around the star.

As the vector phase is opposite for circular polarization states with opposite handedness, the average phase pattern for unpolarized incident light will be flat, and therefore a symmetric PSF without dark holes will be produced.
However, if after the vector-APP a splitting according to circular polarization states is realized, two PSFs with dark holes are obtained.
In order to separate the two circular polarizations and the two corresponding PSFs, a circular polarizing beam-splitter needs to be implemented behind the vector-APP.
As all (practical) polarizing beam-splitter act upon linear polarization states, the two circular polarization states need first be converted into perpendicular polarization directions.
Therefore, a quarter-wave plate (QWP) needs to be positioned after the vector-APP, with its (uniform) fast axis at $\pm 45^\circ$ from the axes of the polarizing beam-splitter, see Fig.~\ref{VAPPprinciple}.
This QWP can be a separate device out of regular crystals, but it could also be implemented as a second liquid crystal layer of the vector-APP, with a uniform orientation of the liquid crystals (which now have a retardance of a quarter-wave).
The vector-APP obviously needs to be inserted into a pupil plane. 
It is most convenient to locate the polarizing beam-splitter right after the vector-APP and the QWP, and therefore a Wollaston prism would be the polarizing beam-splitter of choice.
It creates an angle between beams of perpendicular linear polarization by an amount that is adjustable by design.
Since it is close to a pupil plane, a Wollaston prism makes that the two complementary PSFs are separated in the consecutive image plane by a certain amount.
Thus, one can design a system that covers both PSFs with the same detector.

\subsection{Achromatization}\label{achr}
An achromatic design has been made for the classical APP that is based on the combination of two optical glasses\cite{achromaticAPP}, similar to the construction of an achromatic lens. 
In principle two different glass substrates form a quasi-achromatic phase doublet that provide suppression across a decade of wavelength with negligible impact on performance. 
However, the most appropriate choice of glasses leads to optics that will have large gradients of slope, and for several designs with large contrasts and small inner working angles these will lead to total internal reflections within the optic. 
This is one of the main reasons that alternatives to the current manufacturing paradigm were sought after.

If one is able to create retarder material that is half-wave for all required wavelengths, the resulting vector phase plate is essentially achromatic.
The resulting phase pattern is only determined by the orientation pattern of the liquid crystals.
However, by nature half-wave and quarter-wave retarders are chromatic, mostly with a $1/\lambda$ behavior, which leads to PSF leakage (see Sect.~\ref{tolerances}).
Several methods exist to create ``achromatic retarders''.
A first common method is to arrange at least two elements with different birefringent materials with different dispersions, e.g., with inorganic crystals\cite{Beckers,4QPM2} or with liquid crystals\cite{Schirmer}. 
A version of this approach has also been demonstrated using two sets of LC materials mixed into a single birefringent layer\cite{Parri} to achieve a similar effect.
A second common method is to artfully arrange several elements of the same birefringent material, popularized especially by Pancharatnam\cite{Pancharatnam}.
It applies geometric arguments to design multilayer structures that yield a retardance that approaches the target retardance within some tolerance over a broad wavelength range.
For instance, a three-layer structure for which each individual layer is composed of birefringent material that is half-wave (or quarter-wave) at the centre wavelength $\lambda_0$ can be achromatized by rotating the (local) orientation of each layer.
A three-layer structure has a typical bandwidth of $(1 \pm 0.2)\cdot \lambda_0$, whereas a five-layer structure can go up to $(1 \pm 0.4)\cdot \lambda_0$.
Typical tolerances are $\pm$ 0.05 wave for the retardance and $\pm 2^\circ$ for the fast axis orientation over the specified bandwidth.
Also the effective axis angle varies with wavelength by about 2$^\circ$\cite{Keller2000}.
A Pancharatnam achromatization approach has been applied to the vector vortex coronagraph.\cite{Mawet2, vectorvortextech}
If the LC layers are manufactured separately, issues pertain to their mutual alignment, particularly at the location of the central singularity.\cite{vectorvortextech}

A third more recent method uses multiple twisted LC layers on a single alignment layer and substrate, something we call a multi-layer twisted retarder (MTR) \cite{Komanduri+2012, Komanduri2}. 
Most importantly, subsequent LC layers are directly aligned by previous ones, resulting in simple fabrication requirements and a \emph{monolithic} film, and inherently eliminates the alignment between multiple elements in the second method just described. 
Nearly as important, this birefringent layer can be very easily formed on patterned retardation substrates, and spontaneously aligns with the optical axis set by the alignment layer, as we have shown for louvered quarter-wave plates\cite{Komanduri+2012}.
MTRs are tailorable for retardation spectra nearly arbitrary bandwidths and shapes. 
In our initial work, we show achromatic quarter-wave retarders\cite{Komanduri+2012} and in a related work we show the principle for half-wave retardation in a Pancharatnam-type element (polarization gratings\cite{achromaticpolarizationgrating}). 
While detailed and comparative study of MTR performance is in preparation \cite{Komanduri2}, the basic conclusion is that MTRs with two or three
layers can achieve the same or better performance as the standard three\cite{Pancharatnam}and five layer Pancharatnam style achromatic elements.

Furthermore, the bandwidth can even be increased by combining all aforementioned methods.

\subsection{Tolerancing}\label{tolerances}
In the end, the success of a coronagraph depends heavily on its manufacturability and alignability.
Here, we explore the tolerancing for producing and implementing a vector-APP within a real astronomical instrument.
As the original APP has attained an increase in contrast by two orders of magnitude at the second Airy ring on sky\cite{NaCoAPP2}, we will adopt this contrast level as the tolerance for all pertinent errors.
\begin{itemize}
\item If the \textit{retardance of the vector-APP} is not perfectly half-wave (for certain wavelength range), a certain amount of light from the original PSF will be present, as the circular polarizations are not completely swapped.\cite{Mawet1} 
The amount of leakage goes with $\sin^2(\tfrac{1}{2}\Delta \delta)$, which puts a tolerance on the achromaticity of the half-wave material of $\Delta \delta \lesssim$0.03 wave ($\approx 11^\circ$ of retardance).
It is therefore clear that several layers need to be applied for sufficient achromatization.
Alternatively, the spectral bandwidth can be limited.
This requirement can be relaxed if only one circular polarization direction is present, and therefore one of the two complementary PSFs is sacrificed.
This can be effectuated by a polarizer and additional QWP before the vector-APP.\cite{Mawet1, Mawet2}
\item The \textit{apodizing phase pattern, i.e.~the LC orientation pattern} obviously needs to be as close as possible to the ideal case. 
Tolerancing on this needs to be performed using a Monte Carlo approach, and is left for future work. Also the alignment of putative multi-layers for achromatizing is subject to mutual alignment tolerances.
\item If the \textit{QWP retardance} is not perfectly quarter-wave (for certain wavelength range), the two circular polarization states are not fully converted into the linear polarization directions that are to be split by the polarizing beam-splitter.
Hence the remaining circular polarization components will be spread over both PSFs and pollute both of them.
Note that for this case the pollution is by the opposite PSF, and not by the original PSF.
This leakage also follows $\sin^2(\tfrac{1}{2}\Delta \delta)$, end hence the retardance requirements for the QWP are similar to those for the vector-APP.
\item The \textit{extinction ratio of the polarizing beam-splitter} determines how pure the polarization selection if for each beam.
The general tolerance requirement described above directly translates into a requirement for the extinction ratio to be better than $10^{-2}$, which is common for Wollaston prisms, but requires some attention for beam-splitter cube with dichroic splitter coatings. The chromatism of splitting by a Wollaston prism needs to be small enough such that the broadband image does not get smeared by more than $\sim$0.5 $\lambda/D$.
\item Specific care needs to be taken that no \textit{ghosts} can be present in both D-shaped dark holes at the 1\% level of the PSF core.
\item Finally, the PSFs need to be separated sufficiently, such that the \textit{overlapping outer Airy rings and the PSF scattering wings} are darker than the requirement for the holes.
\end{itemize}

\subsection{Wavefront sensing with a VAPP}

In an adaptive optics system, the optical path to the wavefront sensor is different from the optical path to the science camera, starting from the optical element that splits the incoming wavefronts onto the two different paths. 
The AO system cannot sense and correct the differential aberration between these two beam paths, and this non-common path (NCP) aberration leads to a halo of uncorrected speckles in the science camera focal plane. 
These NCP aberrations evolve in both spatial scale and time, leading to speckle structures that do not average down with increasing integration times.  
These NCP aberrations are modulated by the diffraction halo of the telescope to form speckles, and coronagraphs are designed to suppress this halo and therefore reduce the effects of the NCP aberrations and the resultant speckles.

The vector-APP produces a 360 degree field of view around the central star with the telescope diffraction suppressed in to the first Airy ring, providing a higher signal to noise environment in detecting these residual speckles. 
These speckles are then characterized using a technique such as Phase Sorting Interferometry (PSI\cite{PSI1, PSI2}) to determine the phase and intensity of the speckle with respect to the central Airy core. 
These speckles can either be actively removed with an anti-halo closed loop server, or are used to generate model science PSFs for post-processing subtraction at a later time.

Another advantage of the two PSFs that the vector-APP produces is that the device introduces a known phase diversity in between the two beams, i.e. twice the phase pattern of the APP.
Such a configuration can be used to solve for the phase aberration that is common to the two beams.
The way, the science camera can be used as a wavefront sensor to control a deformable mirror.\cite{keller2012spie}

\subsection{Polarimetry with a VAPP}
With the conceptual design described in the previous section, the vector-APP has the functionality of both a coronagraph as well as a polarizing beam-splitter.
The implementation of polarimetry is therefore a sensible consideration.
In most astronomical cases the prime observable for an imaging polarimeter is linear polarization.
However, the vector-APP plus the beam-splitter only acts upon circular polarization.
To turn it into a regular (i.e.~acting on perpendicular linear polarization directions) polarizing beam-splitter, a second QWP needs to be implemented, before the vector-APP, see Fig.~\ref{VAPPprinciple}.
The order of components then becomes: QWP1, vector-APP, QWP2, Wollaston prism.
With the production techniques described above, the two QWPs can be physically connected to the vector-APP, such that they constitute a single device that is to be inserted into a pupil plane before a Wollaston prism (or any other regular polarizing beam-splitter for that matter).
This shows that most existing astronomical polarimeters could easily be furnished with an APP coronagraph, as long as they have an available pupil plane location before the polarizing beam-splitter.
The polarimetric modulator (often a rotating or switchable half-wave device) would then be located before the vector-APP and, obviously, the beam-splitter.

As the PSFs out of the beam-splitter are designed to be considerably different, the polarimetric analysis with a vector-APP would differ from the normal dual-beam polarimeter situation.
In any case, the object of interest will often be located inside one of the dark D-shaped regions, which means that the polarimetric analysis is necessarily according to a single-beam system.
The major disadvantage of a single-beam polarimeter is that the various observations from which the Stokes parameters are derived, are recorded consecutively.
As the polarized Stokes parameters ($Q$ and $U$ for linear polarization) are obtained after subtracting intensity observations for different orientations of the HWP, the results are very susceptible to seeing and variable sky transparency and background. 
It can be shown that such differential effects are cancelled out to first order once the temporal modulation is combined with the simultaneous information from the two beams out of the beam-splitter.\cite{Semel1993, Tinbergenbook, Bagnulo+2009, SnikKellerreview}
For the center of the stellar PSF, this dual-beam polarimetry can be applied, as long as the AO delivers a high Strehl, which is required anyhow for the coronagraphy to work properly.
This means that the polarimetric correction for a point source can be derived by comparing the results for the single-beam on either side with the results from the full-blown dual-beam demodulation.
Hopefully, this correction can then be applied to point sources inside the dark hole on either side, or to an extended circumstellar structure through deconvolution.

\section{FIRST PROTOTYPE RESULTS}

\subsection{Production}

\begin{figure}[t]
\centering
\includegraphics[width=\textwidth]{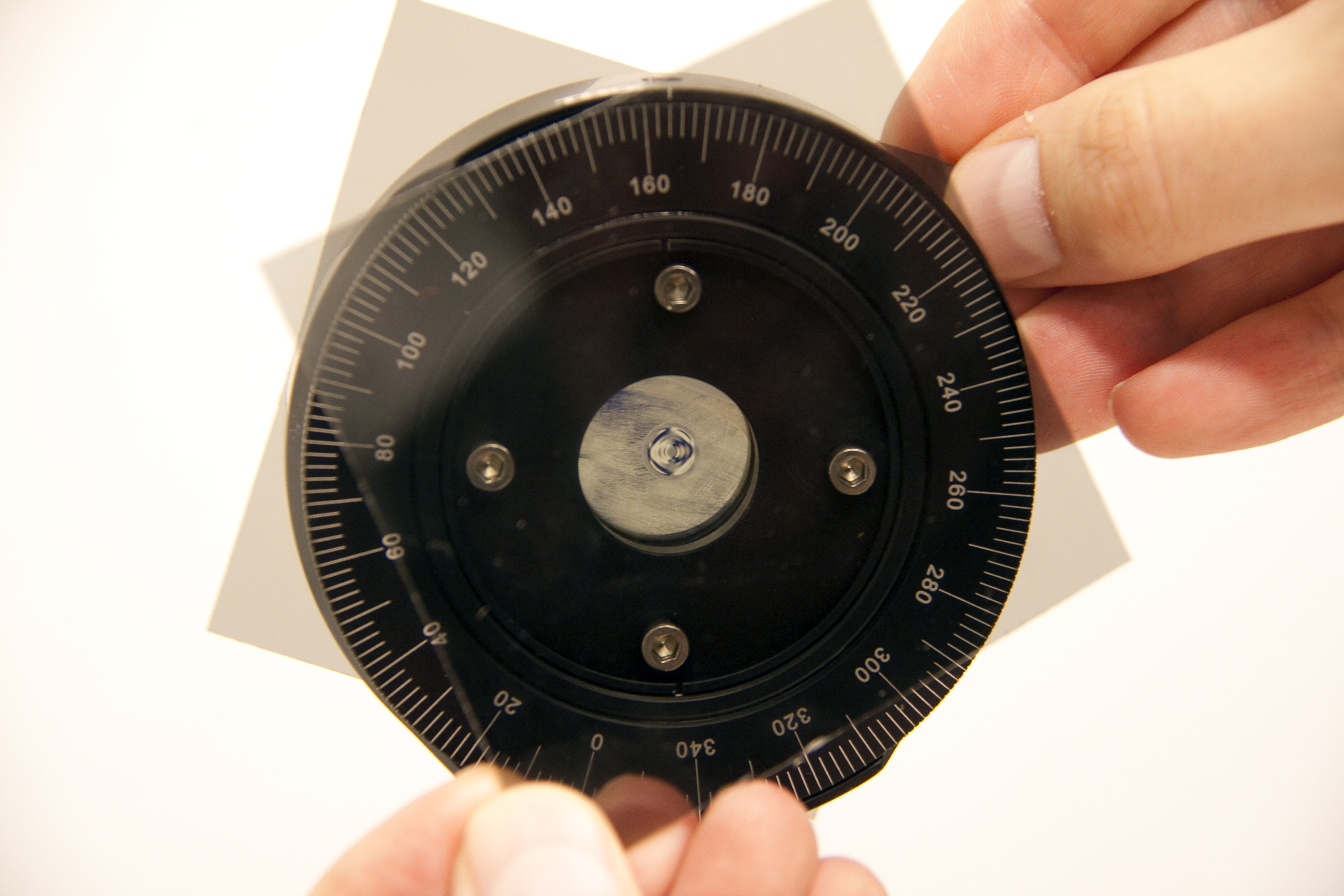}
\caption{The vector-APP prototype in between polarizers. The structured region the the center is to be inserted into the pupil plane of a high-contrast imaging system.}
\label{VAPPphoto}
\end{figure}

Fabrication of the vector-APP prototype followed three stages: patterning, coating, and assembly. 
First, a linear photopolymerizing (LPP) polymer (ROP-108, from Rolic) was applied to a glass substrate using standard processes\cite{Schadt3}. 
This LPP was then exposed using a 325 nm HeCd laser, capturing the desired vector-APP orientation $\theta(x,y)$. 
The fluence (or dose) of the exposure was approximately 15 mJ/cm$^2$. 
Second, three layers of a liquid crystal polymer \cite{Broer} (RMS03-001C, $\Delta n = 0.15$ @ 633 nm, from Merck) were spin-coated to bring the half-wave retardance to 633 nm.
Each layer was polymerized after its coating with a high-power UV LED, prior to application the next layer, so that the result was a solid monolithic film adhered to the single glass substrate. 
Third, the sample was laminated with a protective cover glass using a standard optical adhesive (from Norland). 
The RMS wavefront error was measured as $\lambda/4$ @ 633 nm, using a Shack-Hartmann sensor. 
This level of distortion is comparable to other liquid crystal based elements with similar processing, and is largely a by-product of the lamination process and the use of unpolished float-glass.

\begin{figure}[p]
\centering
\includegraphics{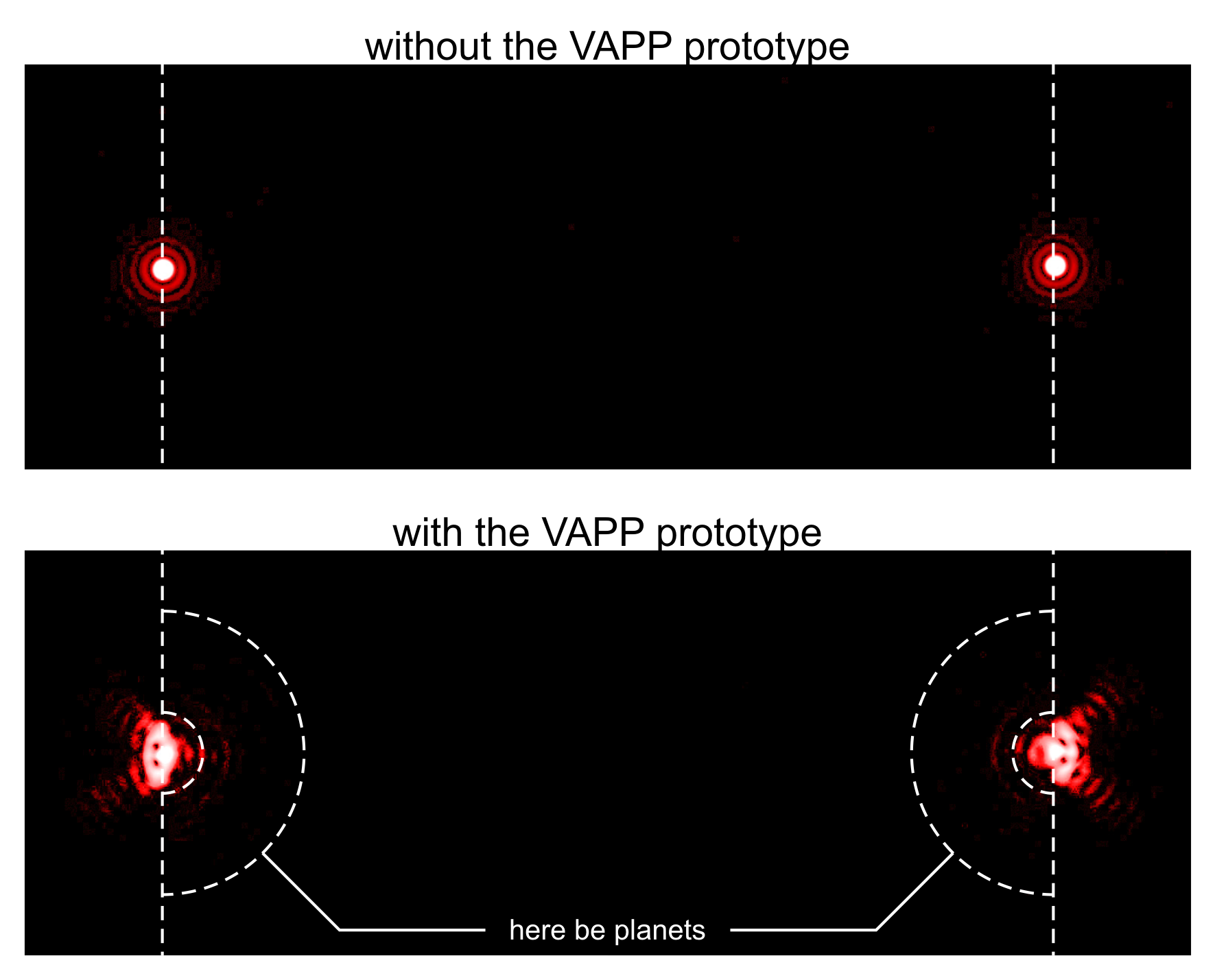}
\caption{First results with the prototype of the vector-APP at 633 nm.}
\label{protoresults}
\end{figure}

\begin{figure}[p]
\centering
\includegraphics[width=0.4\textwidth]{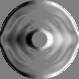}
\hspace{0.05\textwidth}
\includegraphics[width=0.4\textwidth]{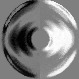}
\caption{a) Ideal APP phase pattern (including phase wrapping as in measurement). b) Measurement of the vector-APP prototype phase pattern.}
\label{protomeas}
\end{figure}

\subsection{Performance}

The prototype vector-APP was optimized for the HeNe 633 nm laser wavelength.
The device was tested at the AO testbed facility at Leiden University.
It was located in a 6-mm pupil plane, and followed by an achromatic (quartz+MgF$_2$) QWP and a quartz Wollaston prism.
The first results are presented in Fig.~\ref{protoresults}.
The two PSFs are clearly antisymmetric and most of the light on the inside of the PSFs has been transported outwards.
It is however also clear that some speckles still exist in the D-shaped areas.
To investigate the causes of this, we investigated the retardance and the orientation pattern of the prototype vector-APP directly by sandwiching it between polarizers and implementing a pupil-imaging lens.
We obtain four frames with different orientations of the polarizers: parallel and crossed with the first polarizer vertical (and parallel to the laser polarization), and parallel and crossed with the first polarizer at 45$\circ$.
The retardance $\delta$ and the orientation pattern $\theta(x,y)$ are then obtained from:
\begin{eqnarray}
\cos(\delta) & = & \left[ \frac{I_{0\parallel}-I_{0\perp}}{I_{0\parallel}+I_{0\perp}} + \frac{I_{45\parallel}-I_{45\perp}}{I_{45\parallel}+I_{45\perp}}  \right] -1\,;\\
\cos(4\theta(x,y)) & = & \frac{1}{2} \left[ \frac{I_{0\parallel}-I_{0\perp}}{I_{0\parallel}+I_{0\perp}} - \frac{I_{45\parallel}-I_{45\perp}}{I_{45\parallel}+I_{45\perp}}  \right] / \sin^2(\delta/2)\,.
\end{eqnarray}

The advantage of this method over rotating the vector-APP or continuously rotating the polarizers is that it is independent of the flat-field and laser polarization (as long as the first polarizer is not crossed with it), and alignment of the four frames is trivial.
However, due to the non-uniqueness of the \textit{arccos} function, not the full range of orientation angles are retrieved and wraparounds occur.
With the geometry as described, the first wraparound occurs around the antisymmetry line of the APP, which makes that the resulting phase pattern measurement gives a symmetric result. 
By changing the sign again around the antisymmetry line, most of the APP phase pattern is measured directly, see Fig.~\ref{protomeas}.
For the largest phases $|\phi|>\pi/2$ more wraparounds occur.
The typical errors on the generated phase pattern are within $\pm5^\circ$ (corresponding to orientation errors of $\pm2.5^\circ$), although there is also a wide tail to the error distribution. 

In the near future we will commission an imaging Mueller matrix ellipsometer that can be used to fully characterize vector-APPs at all visible wavelengths.

\section{CONCLUSIONS \& OUTLOOK}

We conclude from the prototyping that the vector-APP is a promising and practical new coronagraph.
Further development is required to meet the stringent requirements on the production of a vector-APP that yields a contrast of $10^{-4}$ at the location of the second Airy ring.

To better achieve the desired phase/LC orientation profile, we plan to optimize the overall calibration, and increase the spatial resolution, both of which will allow for better reproduction of finer features. 
An achromatic vector-APP will be developed using the MTR principles reported earlier for achromatic waveplates\cite{Komanduri+2012, Komanduri2}, and an achromatic quarter-wave MTR\cite{Komanduri+2012} will be fabricated and attached directly to the vector-APP.

As the production method for the vector-APP allows for the generation of very strong spatial gradients (which cannot be manufactured for a regular phase plate), more extreme solutions for the phase pattern that yield a better coronagraphic performance can be considered.

As the vector-APP is a solid-state device that can be inserted into any filter wheel at the location of a pupil plane, it is easily implemented into several astronomical instruments.
We therefore aim to obtain on-sky results relatively soon.
Several instruments that are optimized for polarimetric high-contrast imaging at visible wavelengths such as ExPo\cite{ExPo} and SPHERE-ZIMPOL\cite{SPHEREZIMPOL1,SPHEREZIMPOL2} already contain a polarizing (cube) beamsplitter, such that only the vector-APP itself and an (attached) QWP needs to be inserted.
Also the AO-assisted imaging infrared polarimeter MMT-POL\cite{MMTPOL} furnishes an excellent platform to test the vector-APP, as it already contains a Wollaston prism that has been optimized for the 1-5 $\mu$m range.
The instrument fully benefits from the telescope's adaptive secondary mirror that provides a PSF with high Strehl ratio and without instrumental polarization.
The CLIO\cite{CLIO} instrument at the MMT shares this benefit, and is therefore also considered a test-bed for the vector-APP.
Future uses of Apodizing Phase Plates include their application on the Large Binocular Telescope in LMIRcam\cite{achromaticAPP}, where the two apertures are combined to form one large coherent aperture.

\acknowledgments     
MNM and MJE gratefully acknowledge the support of the US National Science Foundation (grant ECCS-0955127).


\scriptsize{

\bibliography{Sniketal2012-VAPP-SPIE_8450-21}

}

\end{document}